\documentclass[12pt,preprint]{aastex}

\received{2007 May 24}
\begin{document}

\title{Mass segregation in very young open clusters
      -- A case study of NGC 2244 and NGC 6530}

\author{L. Chen}
\affil{Shanghai Astronomical Observatory, Chinese Academy of
Sciences, 80 Nandan Road, Shanghai 200030, P.R. China}
\email{chenli@shao.ac.cn}

\author{R. de Grijs}
\affil{Department of Physics and Astronomy, The University of
Sheffield, Hicks Building, Hounsfield Road, Sheffield S3 7RH, U.K.;
and National Astronomical Observatories, Chinese Academy of Sciences,
20A Datun Road, Chaoyang District, Beijing 100012, P.R. China}
\email{R.deGrijs@sheffield.ac.uk}

\and

\author{J. L. Zhao}
\affil{Shanghai Astronomical Observatory, Chinese Academy of
Sciences, 80 Nandan Road, Shanghai 200030, P.R. China}

\begin{abstract}
We derive the proper motions, membership probabilities, and
velocity dispersions of stars in the regions of the young ($\sim
2-4$ Myr-old) open clusters NGC 2244 (the central cluster in the
Monoceros R2 association) and NGC 6530 (the dominant cluster in
the Sgr OB1 association) from photographic plate material obtained
at Shanghai Astronomical Observatory, with time baselines of 34
and 87 years, respectively. Both clusters show clear evidence of
mass segregation, but they do not exhibit any significant
velocity-mass (or, equivalently, a velocity-luminosity)
dependence. This provides strong support for the suggestion that
the observed mass segregation is -- at least partially -- due to
the way in which star formation has proceeded in these complex
star-forming regions (``primordial'' mass segregation). Based on
arguments related to the clusters' published initial mass
functions, in conjunction with our new measurements of their
internal velocity dispersions ($\sim 35$ and 8 km s$^{-1}$ for NGC
2244 and NGC 6530, respectively), we provide strong arguments in
favor of the dissolution of NGC 2244 on very short time-scales,
while we speculate that NGC 6530 may be more stable against the
effects of internal two-body relaxation. However, this latter
object may well be destroyed by the strong tidal field prevalent
at its location in the Galactic plane in the direction of the
Galactic Center.
\end{abstract}


\section{Introduction}

Open clusters are considered excellent laboratories for studies of
stellar evolution and the dynamics of stellar systems. A large
fraction of Galactic open clusters are situated close to the Galactic
plane, thus potentially making them appropriate objects for probing
the structure and evolution of the Galactic disc -- although the
presence of dust in the Galactic plane restricts their usefulness in
this sense.

One of many important aspects in open cluster studies is related to
the effects of mass segregation; that is, the more massive member
stars in the cluster will be more centrally concentrated and/or have a
distinct distribution in velocity space compared to the lower-mass
members. The key observational questions of interest in open star
cluster research include whether or not mass segregation exists in a
given star cluster and, if so, what the extent of this effect is in
relation to the cluster's age. For cluster samples as a whole, the
similarities and differences of the effects of mass segregation for
clusters of various ages, abundances and galactocentric distances are
of great interest. All of these properties are related to the
formation process as well as the dynamical evolution of clusters. They
are also (sometimes significantly) influenced by the Galactic tidal
fields, in particular in the outer regions of a given cluster.

Young open clusters are important stellar systems for our
understanding of the star-formation process. In very young
clusters, most of the low-mass stars will still be in their
pre-main-sequence (PMS) evolutionary stage, and one may expect to
derive global stellar initial mass functions (IMFs) that are not
yet significantly affected by the effects of stellar and dynamical
evolution. However, if a significant amount of ``primordial'' mass
segregation (i.e., mass segregation intrinsic to the
star-formation process itself) is present, this would seriously
complicate the interpretation of an observed stellar luminosity
function (LF) at a given position within a star cluster in terms
of its IMF \citep{gri02a,gri02b}.

In this paper, we use photographic plate material with longer time
baselines than published previously, in order to determine the proper
motions and membership probabilities of stars in the regions of two
very young Galactic open clusters NGC 2244 and NGC 6530. We aim to
quantify the possible effects of mass segregation and place these
objects in the context of their expected future evolution.

\subsection{NGC 2244}

NGC 2244 (also referred to as NGC 2239) is the core OB cluster of
the Monoceros OB2 (Mon OB2) association, embedded in the Rosette
Nebula (NGC 2237/2246), a well-developed ``blister'' H{\sc ii}
region. This system is located in the northwest quadrant of the
Rosette Molecular Cloud complex, which itself is one of the most
massive giant molecular cloud (GMC) complexes in the Milky Way
(see \citet{li05b}); such GMC complexes are particularly conducive
to the formation of massive OB stars. Because of the strong
ionizing radiation from the massive OB stars in the core of NGC
2244 (see, e.g., \citet{per87} for a census of the OB stars in the
cluster), the cluster has excavated the GMC complex, allowing us a
deeper look into the cloud than otherwise possible. NGC 2244
represents the youngest of two or three subgroups of OB stars and
stellar aggregates in the Mon OB2 association \citep{li05,li05b}.

Since NGC 2244 is one of the youngest star-forming open clusters
in the Milky Way, and relatively nearby, it has been the subject
of a number of detailed studies. Using the WEBDA open cluster
database \citep{merm03} as our guide, the equatorial coordinates
of the cluster center are $(\alpha,\delta)_{2000} = (06^{\rm
h}31^{\rm m}55^{\rm s}, +04^{\circ}56'30'')$, and the Galactic
coordinates $(l,b) = (206^{\circ}.31, -2^{\circ}.07)$, i.e., the
cluster is located quite close to the direction of the Galactic
anticenter.

\citet{li05}, using near-infrared imaging from the Two Micron
All-Sky Survey (2MASS), estimated a mean visual extinction, $A_V
\simeq 1.5$ mag within a radius of 10 arcmin centered on the
apparent center of the H{\sc ii} region. They pointed out that
this independent low near-infrared extinction estimate was
consistent with previous estimates in the optical, which is also
supported by results from CO and C$^{13}$O maps (e.g.,
\citet{bli86}; \citet{wil95}). In the optical, the most
comprehensive extinction analysis for the cluster was published by
\citet{mas95}, who reported a mean reddening of $E(B-V)=0.48$ mag,
with a non-negligible amount of differential reddening across the
cluster: their reddening values for individual cluster stars range
from $E(B-V)=0.08$ to 0.98 mag. This is fully consistent with the
results of \citet{par02}; and references therein), who found
$\langle E(B-V)\rangle = 0.47 \pm 0.04$ mag from the individual
reddening values for 28 cluster member stars brighter than $V =
14$ mag, and a smaller amount of differential reddening, ranging
from $E(B-V) = 0.40$ to 0.56 mag. They also found that NGC 2244 is
most likely characterized by a ``standard'' reddening law, with a
total-to-selective extinction ratio of $R_V = 3.1 \pm 0.2$. As
such, we adopt a foregound reddening of $E(B-V)=0.48$ mag in this
paper.

The distance to NGC 2244 is reasonably well established, at $1.4
\la D \la 1.7$ kpc, and generally consistent among different
distance indicators used in the literature (see also \citet{par02}
for a review). For instance, \citet{per87} obtained $D = 1670 \pm
125$ pc from the distance moduli of the individual OB-type stars
in the cluster. More recently, \citet{hen00} derived a distance of
$D = 1.39 \pm 0.1$ kpc to the cluster based on spectroscopic
observations of one of the cluster member stars, the eclipsing
binary V578 Mon (HD259135). \citet{par02} derived a distance
modulus of $(m-M)_{V,0} = 11.1$ based on $UBVI$ photometry,
corresponding to $D = 1.7$ kpc. This larger distance value seems
to be favoured by many authors in the field (e.g.,
\citet{ogu81,sab01}), and as such we will adopt it here as well.

For the age of NGC 2244, most authors agree on a value of 2--4 Myr
for the main sequence (MS) turn-off age
(\citet{tur76,ogu81,mar82,per89,per91,mas95,par02} and references
therein). In addition, recently an age estimate of $2.3 \pm 0.2$
Myr was derived from the age of the eclipsing binary V578 Mon
\citep{hen00}. Based on high spatial resolution multi-filter
photometry of NGC 2244, \citet{berg02} concluded that while most
of the cluster stars are aged up to 3 Myr, significantly younger
low-mass stars (5\% or more of the cluster stars) exist in the
cluster as well, indicating that star formation is still in
progress in the Rosette Nebula/NGC 2244 region (see also
\citet{per87,per89,par02}, the latter authors found a MS turn-off
age of 1.9 Myr, but a pre-MS age spread of about 6 Myr). However,
they also note that up to 45\% of the cluster stars may be older
than 3 Myr, thus suggesting an earlier star-forming episode in the
region. Moreover, the dynamical age of the nebula is just $0.2 -
0.6$ Myr \citep{mat67}, while the dynamical age of the H{\sc i}
shell is as old as 4 Myr (see \citet{li05a}).

The first proper motion study of this cluster \citep{sch58}
identified 22 probable members. Subsequently, \citet{mar82}
performed more precise proper motion measurements of 287 stars
down to $V=14$ mag and selected 89 stars with membership
probabilities, $p\geq0.7$, and data quality, $q \ge 0.9$, in an
area of $35\times44$ arcmin$^2$. However, \citet{par02} cautioned
that their ``member'' sample may contain several non-cluster
members of the Mon OB2 association because of the large
rectangular field of view employed by the \citet{mar82} study
compared to the cluster's angular size of $\sim 24$ arcmin and the
roughly spherical shape described by \citet{per91}. Using the
\citet{mar82} astrometric membership determinations, combined with
spectroscopically determined stellar MK classifications of 193
stars by \citet{ver91}, \citet{par02} redetermined the membership
probabilities of the brightest cluster stars (30 OB stars and 21
pre-MS members and candidates brighter than $V=18$ mag).

\citet{sab01} and \citet{hig02} redetermined the cluster's stellar
membership probabilities based on the early astrometric data first
analyzed by \citet{mar82}. They confirmed 92 (out of 202) stars as
physical members of the cluster, with $p > 0.5$. In this paper, we
will analyze high-quality, wide-field photographic plates of
Shanghai Astronomical Observatory, with an maximum likelihood
algorithm involving careful assessment of the distributions of
both the proper motions and the positions, thus allowing us to
arrive at a valuable, independent estimation of membership
probabilities.

\subsection{NGC 6530}

NGC 6530 is another prime example of an extremely young open
cluster that has been the subject of a number of previous studies.
As the core cluster of the Sgr OB1 association, it is projected
onto the eastern part of the Lagoon Nebula (M8), which is one of
the brightest nearby nebulae and H{\sc ii} regions \citep{rau02}.
The Lagoon Nebula is the illuminated part of the GMC from which
the cluster formed by generating a ``blister'' H{\sc ii} region
due to the ionizing radiation from its hot massive stars (e.g.,
\citet{lad76}. Once again, this allows us to probe into the depths
of this GMC region much better than would have been possible
otherwise.

The equatorial coordinates of NGC 6530 are $(\alpha,
\delta)_{2000}=(18^{\rm h}04^{\rm m}31^{\rm s},
-24^{\circ}21'30'')$ and the Galactic coordinates
$(l,b)=(6^{\circ}.083, -1^{\circ}.331)$, i.e., the cluster is
located in the direction not far from that of the Galactic Center.
This implies that one has to be very careful in disentangling
cluster members from field interlopers (see also \citet{sun00}. As
a case in point, in early work on this cluster, \citet{wal57}
obtained photo-electric magnitudes and colors for 118 stars in the
NGC 6530 region and proposed that a band of its low-mass members
above the MS in color-magnitude space may be PMS stars that are
still in the stage of gravitational contraction. It has since been
shown (e.g., \citet{the60,alt72,chi81}) that many of these PMS
candidates are, in fact, foregound or background field stars (see
the discussions in \citet{anc97}, and \citet{sun00}).

Subsequently, the properties of this very young cluster have been
studied extensively by different authors, leading to an age estimate
of 1.5--3.0 Myr \citep{alt72,sag78,sun00}, and the cluster's distance
to the Sun has been estimated to be in the range from 0.6--2.0 kpc
\citep{wal57,alt72,sag78,ste95,sun00,lok01,pri05}.

The most recent median age estimate for the cluster, based on
ground-based $BVI$ photometry down to $V=22$ mag \citep{pri05} is
2.3 Myr, i.e., well inside the range quoted by previous authors,
although the same team also derived a much younger median age of
0.8 Myr (\citet{dam04}; see also \citet{sun00} for the pre-MS
stars; \citet{dam06}). Both estimates are not necessarily in
conflict with each other, because the cluster appears to host
stars formed over a significant length of time. While
\citet{anc97} argued that star formation in the cluster may have
been ongoing since a few $\times 10^7$ ago, until the present,
\citet{sun00} -- based on their detailed study of low-mass PMS
stars -- argue that the time-scale for star formation in NGC 6530
must have been much shorter, starting only a few $\times 10^6$ yr
ago. Nevertheless, this clearly implies that the cluster
environment has been an active star-forming environment for a
significant period (see also \citet{sun00}).

More specifically, \citet{dam06} conclude, based on a detailed
analysis of the distribution of optical/infrared-excess stars
throughout the region, that star formation has proceeded rather
undisturbed and for a longer period in the northern cluster region
than near the cluster center. This is in support of suggestions by
\citet{lad76} and \citet{dam06} that the star-formation activity
may have proceeded from north to south in the cluster region,
which is further supported by a spatial age segregation analysis
by \citet{pri05}.

The most recent distance estimate to NGC 6530, based on $BVI$
photometry, was derived by \citet{pri05}, $D = 1250$ pc,
corresponding to a distance modulus $(m-M)_{V,0} \simeq 10.48$
mag. This is towards the bottom end of the range in possible
distances quoted by a variety of authors, $1300 \la D \la 2000$ pc
(\citet{pri05}, their table 1), generally obtained photometrically
from the brightest cluster member stars. In contrast, it is
significantly larger than the distance obtained from Hipparcos
parallaxes of seven member stars, $D = 560 - 711$ pc
\citep{lok01}. In other recent studies, both \citet{anc97} and
\citet{sun00} determined a best distance estimate to NGC 6530 of
$D = 1.8 \pm 0.1$ kpc [$(m-M)_{V,0} = 11.25 \pm 0.1$ mag, based on
optical photometry and a detailed consideration of the extinction
across the cluster (see also \citet{mcc90}). In this paper, we opt
to use the \citet{pri05} distance determination, $D = 1250$ pc.

Following the WEBDA compilation of cluster parameters, we adopt a mean
reddening $E(B-V)=0.35$ mag (\citet{sun00}; see also \citet{chi81}),
in addition to a foreground reddening of $E(B-V)=0.17$ to 0.20 mag
\citep{mcc90,anc97,sun00,pri05}. There is significant variation in
reddening across the NGC 6530 region, as is evident from both the
distribution of stars in color-magnitude space and from the spatial
variation of stellar density (\citet{dam04}, and references
therein). In fact, the amount of differential reddening ranges from
$E(B-V)=0.25$ to 0.50 mag across the cluster (\citet{sun00}; also
\citet{sag78}, but see \citet{anc97}), although \citet{dam04} also
uncovered evidence for the presence of a large number of heavily
reddened objects, characterized by $A_V \la 20$ mag.

\citet{alt72} were the first to determine the membership
probabilities of 363 stars in a $60'\times32'$ region centered on
NGC 6530 from relative proper motion data with a time baseline of
34 years. They found 76 stars with membership probabilities, $p$,
greater than 0.5 with $V \le 13.6$ mag. More recent studies
resulted in firm membership determinations for 451 stars with $p
\ge 0.9$ \citep{zha06}, and 237 stars \citep{pri07}, the latter
study based on spectroscopic observations around the lithium and
H$\alpha$ lines. \citet{dam06} has suggested that the cluster may
contain $\ge 1100$ members.

It should be noted that a reliable stellar membership determination is
crucial for resolved star cluster studies. The Shanghai Astronomical
Observatory has in its archives wide-field photographic plates with a
long time baseline of these two open clusters, thus providing
valuable, independent observational materials for high-accuracy
measurements of stellar proper motions, and for the determination of
cluster membership. This constitutes a strong starting point for
further detailed cluster studies, including those dealing with the
effects of mass segregation, as we show in this paper.

In this paper, we use photographic plate material with longer time
baselines or higher quality than published previously, in order to
determine the proper motions and membership probabilities of stars
in the regions of both clusters (Sect. \ref{obs.sec}). We present
a new, detailed investigation of these two clusters based on our
membership determination (Sect. \ref{members.sec}), of their
stellar luminosity functions and the observed effects of mass
segregation (Sect. \ref{results.sec}). In Sect.
\ref{discussion.sec} we discuss the clusters in the context of
their evolutionary state and summarize the main results obtained
in this paper.

\section{Observations}
\label{obs.sec}

For the open clusters NGC 2244 and NGC 6530, all photographic plates
used for the proper motion reduction were taken by the 40-cm refractor
telescope (with a focal length of 6895 mm) at the Sheshan
(Z\^{o}--S\`{e}) Station of the Shanghai Astronomical Observatory,
operated by the Chinese Academy of Sciences. For NGC 2244, four plates
for each of the first and second epochs were taken in 1963--1964, and
1998--1999, respectively, spanning a time baseline of 35 years. For
NGC 6530, two first epoch plates and four second plates were taken in
1912 and 1999, respectively, spanning 87 years. The field of view
achieved is $2^{\circ}\times2.5^{\circ}$ and
$1.5^{\circ}\times1.5^{\circ}$, for the first and second epoch plates,
respectively. No filter was used to obtain these plates; their
sensitive wavelength region is close to the $B$ band and the 30 min
exposure time resulted in a brightness limit of about 16 mag.

The stellar coordinates on the plates of both clusters were
measured with the Photometrics Data Systems (PDS)
microdensitometer \citep{stil92} at the Dominion Astrophysical
Observatory (DAO) in Canada. For each plate, raster scanning was
performed over the full $60\arcmin \times 60\arcmin$ central
regions of the clusters. The scanning diaphragm used had a size of
$20.5\mu$m squared, with both a step length and line separation of
$20\mu$m; a speed of 15 mm s$^{-1}$ was used. In the region where
we performed the full scan, stellar images were extracted and
their rectangular coordinates, $(x, y)$, obtained using the
processing software provided by DAO. After the preliminary
identification, the measured coordinates on each plate were
transformed linearly to a unified system using six plate
constants, and those with residual errors $\ge 15\mu$m were
excluded. The cross-identification resulted in a final set of 498
stars for NGC 2244 and 365 stars for NGC 6530, respectively, with
each star appearing on at least one plate of both the early {\it
and} later epochs.

\section{Proper motion reduction and membership estimates}
\label{members.sec}

In the reduction of our proper motion data, an iterative
central-overlap technique \citep{rus76,wan96} was used. For NGC 2244,
78 Tycho-2 stars were selected as reference stars; their reference
positions and proper motions were adopted based on the Tycho-2
Catalogue \citep{hog00}. During the iterative solution, three stars
with extraordinarily large residual errors were removed. The final
solution provided us with robust position and proper motion data of
495 stars in the cluster region. Similarly, 79 Tycho-2 catalogue stars
were used as reference stars for NGC 6530, and only 1 star with an
extremely large residual error in the iterative solution was
removed. This resulted in a total of 364 stars with final position and
proper motion values. The internal levels of uncertainty on the
position and proper motion of the individual stars are given in Table
1.

Estimating the membership probability of the observed cluster
stars involves a careful assessment of the distributions of both
the proper motions and the positions. We assume that in position
space the surface number density of the field stars follows a
uniform distribution, $\psi_{\rm f}$, while the cluster members
approximately follow a Gaussian profile (see, e.g., \citet{wen06},
for a preliminary analysis of the stars in the NGC 6530 region
with membership probabilities $p \ge 0.9$), $\psi_{\rm c}$, with a
central density, $n_{0}$, and a dispersion, $\alpha$. Then,
\begin{eqnarray}
\psi_{\rm f} = n_{\rm f}, &
\psi_{\rm c}(r_{i})=n_{0}\exp(-\frac{r_{i}^{2}}{2\alpha^{2}} ) \quad , \nonumber
\end{eqnarray}
where $r_i$ denotes the distance of the $i^{\rm th}$ star to the
cluster center. Let $\Psi_{\rm f}$ and $\Psi_{\rm c}$ be the
normalized $\psi_{\rm f}$ and $\psi_{\rm c}$, respectively, and
\(g=n_{0}/n_{\rm f}\). Then we have \(\Psi_{\rm f} + \Psi_{\rm c} =
1\) (by definition), and
\begin{mathletters}
\begin{eqnarray}
\Psi_{\rm f}(r_i)=\frac{1}{1+g
\exp^{-1}(-\frac{r_{i}^{2}}{2\sigma_{\alpha}^{2}})} ,\\
\Psi_c(r_i)=\frac{1}{1+g^{-1}
\exp(-\frac{r_{i}^{2}}{2\sigma_{\alpha}^{2}})}
\end{eqnarray}
\end{mathletters}

Both the field stars and the cluster members have independent proper
motion and position distributions. We assume that the field star
proper motion is characterized by an elliptical Gaussian distribution,
and that of the cluster stars is a circular Gaussian function. The
overall distribution function for the field stars, $\Phi_{\rm f}$, and
for the cluster stars, $\Phi_{\rm c}$, can then be written as:
\begin{eqnarray}
   \Phi_{\rm f} &=& \Psi_{\rm
      f}(r_{i})\times\frac{1}{2\pi(1-\rho^{2})^{1/2}(\sigma_{x0}^{2}+\varepsilon_{xi}^{2})^{1/2}(\sigma_{y0}^{2}+\varepsilon_{yi}^{2})^{1/2}}\nonumber\\
      &
      &\times\exp\{-\frac{1}{2(1-\rho^{2})}[\frac{(\mu_{xi}-\mu_{x{\rm
      f}})^2}{\sigma_{x0}^2+\varepsilon_{xi}^2} \nonumber\\&
      &-\frac{2\rho(\mu_{xi}-\mu_{x{\rm f}})(\mu_{xi}-\mu_{x{\rm f}})}
      {(\sigma_{x0}^{2}+\epsilon_{xi}^{2})^{1/2}(\sigma_{y0}^{2}+\epsilon_{yi}^{2})^{1/2}}
      +\frac{(\mu_{yi}-\mu_{y{\rm
      f}})^2}{\sigma_{y0}^2+\varepsilon_{yi}^2}]\},
\end{eqnarray}
and
\begin{eqnarray}
    \Phi_{\rm c} &=& \Psi_{\rm c}(r_{i})\times
    \frac{1}{2\pi(\sigma_{0}^{2}+\varepsilon_{xi}^{2})^{1/2}(\sigma_{0}^{2}+\varepsilon_{yi}^{2})^{1/2}}\nonumber\\
    & & \times\exp\{-\frac{1}{2}[\frac{(\mu_{xi}-\mu_{x{\rm
    c}})^2}{\sigma_{0}^2+\varepsilon_{xi}^{2}}+
    \frac{(\mu_{yi}-\mu_{y{\rm
    c}})^2}{\sigma_{0}^2+\varepsilon_{yi}^{2}}]\}
\end{eqnarray}

Here, $(\mu_{xi},\mu_{yi})$ is the reduced proper motion of the
$i^{\rm th}$ star and $(\varepsilon_{xi}, \varepsilon_{yi})$ are the
estimated observational errors. $(\mu_{x{\rm f}},\mu_{y{\rm f}})$ and
$(\mu_{x{\rm c}},\mu_{y{\rm c}})$ are the distribution centers of the
field stars and cluster members in the vector point diagram (VPD),
respectively, whereas $(\sigma_{x0},\sigma_{y0})$ and $\sigma_{0}$ are
the intrinsic proper motion dispersions of the field stars and the
cluster members respectively, and $\rho$ is the correlation
coefficient. Together with the parameters $\alpha$ and $g$, there are
10 parameters awaiting determination. These parameters were estimated
by means of a maximum likelihood method. When searching for the
maximum of the likelihood function, a bipartition algorithm was
adopted \citep{wan97}, and the (internal) uncertainty of the parameter
estimation was evaluated based on the second derivatives of the
likelihood function \citep{zha87}. In order to get a handle on the
external uncertainty of our proper motion determinations, we compared
our derived proper motion values to those derived from the Tycho-2
Catalogue, for the 78 stars in NGC 2244 and the 79 stars in NGC 6530
in common in both data sets. This resulted in a mean deviation of
$\sim 2$ mas yr$^{-1}$, with a dispersion of $\sim 3$ mas yr$^{-1}$,
in very good agreement with the internal level of accuracy of the
Tycho-2 Catalogue.

After the distribution parameters are determined, the membership
probability of the $i^{\rm th}$ star can be calculated as:
\begin{equation}
p_i=\frac{\Phi_{\rm c}(i)}{\Phi(i)}=\frac{\Phi_{\rm c}(i)}{\Phi_{\rm
f}(i)+\Phi_{\rm c}(i)}
\end{equation}

Table 2 gives the maximum likelihood estimates of the distribution
parameters, and the corresponding uncertainties, for both
clusters. We used the distribution parameters to calculate the
membership probabilities of the individual stars in the NGC 2244
and NGC 6530 regions. Figure 1 shows the VPD for both NGC 2244 and
NGC 6530 (see also the preliminary analysis \footnote{We note that
whereas \citet{wen06} published a preliminary analysis based on
the photographic plates also used in the present paper, we
emphasize that the results in the present paper are based on a
thorough and significantly improved re-analysis of the entire data
set (e.g., in terms of the statistical treatment of the
uncertainties and membership probabilities, as well as the
determination of the basic cluster parameters).} by \citet{wen06};
the solid points represent the stars with membership
probabilities, $p \ge 0.9$, whereas the crosses indicate the $p
\le 0.5$. The open circles have $p$ values in between these
boundaries. The VPDs already show that our membership
determination appears to be quite effective. To underscore this
point, Fig. 2 shows the histograms of the membership
probabilities; the left-hand panel is for NGC 2244 and the
right-hand panel for NGC 6530. Returning to the clusters' VPDs,
both \citet{sab01} and \citet{hig02} published VPDs for NGC 2244
based on independent proper motion analyses. Although their
samples of cluster members are much smaller than ours (of order
100 member stars), the VPDs are qualitatively similar in extent as
well as morphology. We note, however, that in particular the
\citet{hig02} VPD appears somewhat more elongated in the $y$
direction than our VPD in Fig. 1. We will return to this issue
below.

From Fig. 2, one can deduce {\it quantitatively} that the
discrimination of membership for NGC 2244 is quite effective. Of
the total of 495 stars, about 16\%, or 78 stars, have
probabilities in the range of $0.3<p<0.7$. As an aside, we note
that a comparison of the membership histograms in \citet{sab01}
and Fig. 2 shows that our membership determination is more
efficient than theirs. Meanwhile, for NGC 6530, the separation of
members from the field stars is even more effective, with only a
few stars having probabilities around $p=0.5$, i.e., only 4.4\%
(or 16 stars) have $0.3<p<0.7$; 451 stars have $p \ge 0.9$ (see
also \citet{zha06} for a discussion of the results in relation to
the earlier work by \citet{alt72}). We note that in the context of
NGC 6530, the \citet{pri05,pri07} and \citet{dam06} studies cover
a much smaller field of view (diameter $\sim 20$ arcmin) than our
observations presented here (Sect. \ref{obs.sec}), although at a
higher spatial resolution. Secondly, the \citet{pri07} sample,
which -- in essence -- only covers the cluster core, is fainter
($B \ga 15$ mag) than ours ($B \la 14$ mag). As such there is
virtually no overlap between both data sets.

\section{Luminosity functions and the effects of mass segregation}
\label{results.sec}

\subsection{Star counts and the cluster sizes}

Star counts allow us to statistically determine the properties of
clusters with respect to the surrounding stellar background. The
cluster radius itself is one of the most important cluster
parameters, and particularly useful for dynamical studies. From
our results, the member stars of both NGC 2244 and NGC 6530 show
an obvious concentration on the sky, while this is clearly not the
case for field stars. In essence, this supports the result for NGC
2244 by \citet{sab01} based on the cluster's VPD. They determined
robust membership probabilities for 92 stars in the area using
both the cluster's VPD and its membership probability histogram,
akin to our Fig. 2. In the VPD they clearly showed the
(elliptical) distribution of the proper motions of the cluster
members around a centroid, and the very small dispersion of stars
with $p \ga 0.7$, clearly setting them apart from the field stars,
with $p \la 0.6$ (see also \citet{hig02}). As regards NGC 6530,
\citet{wen06} and \citet{zha06} provided compelling evidence for
the superiority of the membership determinations in the present
data set compared to that presented by \citet{alt72}.

In order to derive the radial stellar surface density, we first
chose the cluster center as the median coordinates of all the
member stars with $p\geq0.9$. The adopted centers are at the
(J2000.0) equatorial coordinates of $(06^{\rm h}32^{\rm m}04^{\rm
s},+4^{\circ}55'00'')$ for NGC 2244 and $(18^{\rm h}04^{\rm
m}24^{\rm s},-24^{\circ}21'12'')$ for NGC 6530, respectively.
These updated center positions are slightly different from the
values provided by WEBDA, although the displacements between our
new center coordinates and those provided by WEBDA are well inside
the respective cluster core radii (see below). For comparison,
\citet{li05} determined the 2MASS center position of NGC 2244 as
being very close to the centroid determined from our proper
motions, resulting in J2000.0 equatorial coordinates of $(06^{\rm
h}31^{\rm m}59.9^{\rm s},+4^{\circ}55'36'')$. The radial number
density profile, shown in Fig. 3, was constructed by performing
star counts inside increasing concentric annuli around the cluster
center and normalized by the surface areas covered by the
respective annuli.

Figure 3 shows the radial surface number density profiles of the
cluster members $(p\geq0.9)$ and the field stars $(p\leq0.1)$ in
our observational fields of view, on the left-hand side for NGC
2244 and on the right-hand side for NGC 6530. In each panel, the
profiles for the cluster members and the field stars are
completely different, with the field star distributions
essentially flat, whereas the cluster members exhibit obvious
concentrations towards the cluster centers. This result shows once
again that our membership determinations are reasonably robust and
reliable. The basic structural parameters were derived by fitting
the two-parameter surface density \citet{kin66} profile to our
observational data in a least-squares sense. We used fitting radii
of 32 and 33 arcmin for NGC 2244 and NGC 6530, respectively. The
core radii thus obtained are $9.43 \pm 1.48$ arcmin for NGC 2244
and $4.29 \pm 0.90$ arcmin for NGC 6530, corresponding to $4.7 \pm
0.7$ and $1.6 \pm 0.3$ pc, respectively. Meanwhile, the
half-number radii (i.e., the radii on either side of which half of
the member stars are located) of the clusters were also
determined, as 14 arcmin for NGC 2244 and 21 arcmin for NGC 6530,
or $\sim 6.9$ and $\sim 7.6$ pc, respectively. Our main reason for
using the clusters' half-number radii instead of their half-light
radii is practical: the position and proper motion information,
eventually leading to membership probabilities, is based on the
results from the PDS scans. During this process, some of the
brightest stars had to be avoided because of saturation effects on
the photographic plates (which would lead to unacceptably large
positional uncertainties for our analysis). In other words, our
sample of cluster members is incomplete at bright magnitudes, and
the membership probabilities of the (few) brightest stars are
unknown. The half-number radii are not significantly affected by
the omission of these few stars, which would have an
extraordinarily large contribution to the clusters' total fluxes,
however.

Although \citet{li05} found that the radial profile in the
near-infrared of the NGC 2244 member stars disappears into the
background field star population at a radius of about 20 arcmin
\footnote{See also \citet{li05b}, who claim that the total extent
of the cluster is about 20 pc, which -- at a distance of 1.4 kpc
-- translates into a linear diameter of about 49 arcmin. Similar
size estimates were obtained in the optical by \citet{tow03}, and
in X-rays by \citet{berg02} and \citet{che04}.}, our radial
profile in Fig. 3 clearly shows that we can follow the cluster's
profile out to almost 30 arcmin. This is in keeping with the
cluster radius of 11.8 pc derived by \citet{sab01}, which -- at an
assumed distance of 1.4 kpc -- corresponds to about 29 arcmin. Our
size determination for NGC 6530 is less clear-cut, as Fig. 3
shows: the radial profile composed by its member stars appears to
merge with the field star population at a radius of around 20
arcmin, although this value is rather uncertain. We believe,
therefore, that our result is in reasonable agreement with the
diameter of roughly 35 arcmin quoted by \citet{anc97}.

For both clusters, the model profiles tend to roughly follow the
observational data points in the outer regions, but not in detail.
This suggests that both clusters do not have well-defined outer
radii, but are instead characterized by a significant amount of
substructure. There is, in fact, strong evidence that NGC 2244
possesses distinct substructures. \citet{li05}, using 2MASS
near-infrared imaging, found that the cluster is resolved into an
extended distribution of stellar sources associated with its
compact core in the south center of the blister H{\sc ii} region,
but also into a clear elongated density enhancement in the west.
NGC 6530 also exhibit distinct substructures: in a region towards
the northwest of the actual cluster center there appears to be a
density enhancement of slightly older stars than in the rest of
the cluster (e.g., \citet{dam06}). In addition, the VPDs of both
clusters (Fig. 1; see also \citet{hig02}, in particular) appear to
have axial ratios significantly deviating from unity.

Finally, \citet{li05} and \citet{li05b} point out that instead of
using a \citet{kin66} profile, a more appropriate radial density
profile might be represented by a $R^{-1}$ profile, following
suggestions by \citet{bab04} regarding the radial density
distributions of embedded star clusters. In Fig. 3, we show the
best-fitting $R^{-1}$ profiles for both clusters. For NGC 2244,
the best-fitting profile follows $n(R) = 0.046 + 0.615 R^{-1}$,
while the equivalent profile for NGC 6530 is $n(R) = 0.020 + 0.791
R^{-1}$. We note that whereas the fit to the NGC 6530 radial
profile is statistically good, the NGC 2244 distribution is better
approximated by a King profile.

\subsection{Luminosity distributions}

\citet{mas95} provided their $UBV$ observations centered on NGC 2244
with the KPNO 0.9m telescope and a Tektronix 2048$\times$2048 CCD,
covering a 2800 arcmin$^2$ area. This resulted in accurate photometric
data for a total of 773 stars in the cluster region, down to
$B\approx18$ mag. By matching our photographic plate observations to
the \citet{mas95} sample using a conservative matching tolerance of
0.3 arcsec, 323 stars in common were identified. Furthermore, using a
quadratic function, our initial instrumental magnitudes, $B_p$, were
converted to the $B$ magnitudes in the \citet{mas95} system.

As regards NGC 6530, \citet{pri05} performed $BVI$ observations of the
NGC 6530 region using the ESO 2.2m telescope at La Silla,
characterized by a field of view of $34\times33$ arcmin$^2$. The
limiting magnitude of their work is $V\approx17.5$ mag. We compared
our $B_p$ instrumental magnitudes with the $B$ magnitudes of 62 stars
in common with \citet{pri05}'s sample. Except for a few stars, we
found an obvious linear correlation between the magnitudes, which we
used to convert our $B_p$ photometry to the B magnitudes in the $UBV$
system.

Figure 4 shows the comparison between our instrumental $B_p$
photometry and the $B$ magnitudes of $UBV$ system, for both
clusters. The photometric uncertainties in both our own data and those
taken from the literature are small: \citet{mas95} state that their
uncertainties are $< 0.02$ mag for stars brighter than $V=14.5$ mag,
and ``do not become appreciable until $V=16$ mag''. The photometric
uncertainties in the \citet{pri05} data have a mean value of 0.02 mag,
with a maximum error $< 0.2$ mag.

For member stars in a given open cluster, all at roughly the same
distance, the range in visual magnitudes corresponds to that in
luminosity. In addition, when only main-sequence stars are
considered, the stellar luminosity is a proxy for the stellar
mass. In order to assess the effects of mass segregation, we first
examined the luminosity functions of the member stars in different
(radial) ranges of the clusters. Figures 5 and 6 show the $M_B$
luminosity functions of stars with membership probabilities $p \ge
0.9$ in the two cluster regions.

Figures 5a (top) and b (middle) show the luminosity functions of the
NGC 2244 cluster members in the inner $(R\leq R_{\rm h})$ and outer
$(R> R_{\rm h})$ areas of NGC 2244, where $R_{\rm h}=14$ arcmin is the
half-number radius derived from our full sample of 216 member
stars. Fig. 5c (bottom) shows the cumulative distributions for both
radial regions.

Figures 6a (top), b (middle), and c (bottom) show the equivalent
results for NGC 6530 (see also \citet{zha06}). Here, we derived a
half-number radius of $R_{\rm h}=21$ arcmin from our total sample of
250 member stars.

From these two figures, the effects of mass segregation are
evident: the relative fraction of brighter stars (or,
equivalently, more massive stars) is higher in the inner than in
the outer regions. As an alternative approach to investigate these
mass segregation effects, we calculated the half-number radii for
member stars in different absolute magnitude ranges, i.e., for
$M_{B}\leq 0$ and $M_{B}> 0$ mag. The results of this exercise are
listed in Table 5 (see \citet{zha06} for a complementary
discussion related to NGC 6530; in particular their table 6). It
can be seen that for both clusters, the brighter members are
characterized by smaller half-number radii -- another indication
of the existence of mass segregation in these clusters.

To further examine the mass segregation effects, we also investigated
the radial density distributions of member stars ($p\geq0.9$) in
different luminosity ranges in the clusters, which are shown in Figs.
7 (for NGC 2244) and 8 (for NGC 6530). In each cluster, a
position-dependent LF is found. In essence, the fainter (i.e.,
lower-mass) stars are spread throughout the entire cluster in both
objects, while the brighter (massive) stars are predominantly located
in the inner region.

\subsection{Velocity-luminosity relation}

If the mass segregation effects solely result from the process of
two-body relaxation and the associated energy equipartition, one
should find some evidence for this effect not only in the spatial
distribution of the cluster members, but also in their velocity
distribution: the more massive members will have a smaller velocity
dispersion than the less massive stars.

Accordingly, for true cluster members (with $p\geq0.95$) in
different magnitude ranges, the intrinsic proper motion
dispersions $\sigma_{\rm int}$ as derived from the observed
dispersions $\sigma_{\rm obs}$, and the mean errors intrinsic to
the stellar proper motions were investigated. The final results
are shown in Figs. 9 and 10. It can be seen from these figures
that the intrinsic proper motion dispersions are not significantly
different for various magnitude groups. Thus, our data do not
support any significant velocity (dispersion)-mass correlation,
neither in NGC 2244, nor in NGC 6530 (see also \citet{zha06}).

\section{Discussion and Summary}
\label{discussion.sec}

The traditional explanation for (position dependent) mass
segregation in an open cluster is that it results from a situation
in which all cluster stars approach energy equipartition. However,
this is unlikely the case for NGC 2244 and NGC 6530, since both
clusters are very young, low-density objects (but see
\citet{por07} for counterarguments). With ages of only around 2--3
Myr, their ages are comparable to the time-scale of about only one
or two crossing times.

Theoretically, the nature and degree of ``primordial'' mass
segregation, i.e., mass segregation intrinsic to the
star-formation process itself, is presumably determined by the
properties of the interactions of protostellar material during the
star-forming episode in a cluster. More massive stars are subject
to more mergers, hence accrete even more mass (cf.
\citet{lar91,bon01a,bon01b} and references therein), and therefore
dissipate more kinetic energy. In addition, they tend to form near
the cluster center, in the highest-density region, where the
encounter-rate is highest
\citep{lar91,bon97,bon98b,bon01a,bon01b,bon98a}. This will lead to
an observed position-dependent stellar mass function containing
more low-mass stars at larger radii compared to the mass function
in the cluster center (although low-mass stars are still present
at small radii, see also the discussions in \citet{gri02a,gri02b};
but see \citet{por07} for arguments in favor of excluding
primordial mass segregation to explain recent observations of the
Arches cluster near the Galactic Center).

From an observational point of view, the study of very young star
clusters still embedded in the molecular clouds from which they
originated might give us a handle to constrain the degree of
primordial mass segregation. In the Milky Way, in three such young
star clusters mass segregation effects have been studied in great
detail in the past two decades.

While \citet{lad91} suggested, based on ground-based observations,
that the brighter stars in NGC 2024 (and the Mon R2 complex at
large) seem to be more centrally concentrated than the fainter
cluster members, this evidence was deemed inconclusive by
\citet{car97}. They argued that this result was based on an
incomplete sample of cluster members, although mass segregation
might be limited to the massive, bright OB stars forming in the
very center. These same authors argued that for masses below $2
M_\odot$, mass segregation effects in the Mon R2 complex amount to
only a $\sim 2 \sigma$ result.

Secondly, a combination of both ground-based (e.g., \citet{hil97}
and {\sl Hubble Space Telescope (HST)} observations (e.g.,
\citet{hil98}) of the Orion Nebula Cluster (ONC), and in
particular of its very core, the Trapezium stars, have presented
clear evidence for mass segregation for the $m > 5 M_\odot$
component, with some evidence for general mass segregation down to
$m \simeq 1$--$2 M_\odot$ (\citet{hil98}; see also the review by
\citet{lar93}).

Finally, R136, the central cluster in 30 Doradus in the Large
Magellanic Cloud (LMC; age $\lesssim$ 3--4 Myr; cf.
\citet{hun95}), has been studied extensively, both from the ground
and with the {\sl HST}. A variety of techniques have revealed a
significant overabundance of high-mass stars in its very center,
thus supporting strong mass segregation (e.g.,
\citet{cam92,lar93,mal94,bra96}).

Thus, in most of the (still partially embedded) young star clusters
that can be resolved in individual stars, mass segregation effects are
observed, although to varying degrees. This underlines the importance
of our understanding of the physical processes involved in the
formation and evolution of star clusters, and in particular of the
IMF, which will ultimately determine the time-scale on which a young
star cluster will eventually be destroyed.

We may conclude from the above discussion that the observed
spatial mass segregation in the young open clusters NGC 2244 and
NGC 6530 might have resulted from a combination of both initial
conditions in the early stages of evolution and, possibly,
two-body relaxation process. The latter process will lead to a
manifestation of cluster-wide mass segregation on the half-mass
relaxation time-scale. However, it will proceed much faster in
higher-density regions in the cluster, and among the more massive
stars -- both of these conditions favor an decreased time-scale in
the cluster center.

Finally, we will consider the long-term fate of the two young clusters
discussed in this paper in relation to their (observed) IMF and
velocity dispersion. Both clusters have been scrutinized with the aim
of determining their stellar make-up. Because of their very young
ages, the stellar content in essence reflects the mass distribution at
the time of their birth, as insufficient time has passed for any
significant number of stars to have undergone a full cycle of stellar
evolution (and neither for significant dynamical evolution to have
occurred, as argued above).

\citet{par02} determined the IMF of NGC 2244 in the mass range
$-0.5 \le \log (m_\ast / M_\odot) \le 2.0$. They found a flat IMF
slope, with $\Gamma = -0.7 \pm 0.1$, where the equivalent slope of
the \citet{sal55} IMF would be $\Gamma = -1.35$. Their result is
in close agreement both with that of \citet{mas95} in the range
from $\sim 7$ to $15 M_\odot$, and with that of \citet{per91} for
stars with masses in excess of $4 M_\odot$, $\Gamma = -0.7$.

Such a flat IMF implies that the cluster's stellar population contains
too many massive stars with respect to their lower-mass counterparts
for it to survive the dynamical effects leading to cluster dissolution
for any significant amount of time (see, e.g., \citet{smi01} in the
context of the young massive cluster M82-F; and
\citet{che87,che90,goo97} for theoretical arguments in support of this
notion). The precise dissolution time-scale depends sensitively on the
IMF mass range (Kouwenhoven et al., in prep.).  This conclusion is
supported by arguments related to the cluster's velocity
dispersion. Its intrinsic overall velocity dispersion is $\sigma_{\rm
tot}^{\rm int} = 4.45 \pm 0.15$ mas yr$^{-1}$; at the adopted
distance, $D = 1.7$ kpc, this is equivalent to $\sigma_{\rm tot}^{\rm
int} \simeq 35$ km s$^{-1}$. \footnote{We note that if we assume that
this is the cluster's expansion velocity, the dynamical age of the
star-forming region centered on the cluster is of order 0.4 Myr. This
is remarkably close to the dynamical age of the nebula, estimated at
$0.2 - 0.6$ Myr by \citet{mat67}.} This large linear velocity
dispersion is significantly in excess of the ``typical'' velocity
dispersions older open clusters in the Milky Way characterized by a
Salpeter-type IMF, or equivalent (de Grijs et al., in prep.).  Thus,
the combination of these arguments leads us to conclude that NGC 2244
will most likely dissolve on a short time-scale (see also
\citet{li05b} for a more conservative back-of-the-envelope calculation
supporting our conclusion). -- if not due to internal two-body
relaxation effects, then most likely because of external shocks
expected to operate near its location in, or close to, the Galactic
disk (although we note that these effects may not be very significant
in the Galactic anticenter direction)

NGC 6530, on the other hand, exhibits a power-law IMF slope of
$\Gamma = -1.22 \pm 0.17$ for stellar masses between 0.6 and $4
M_\odot$ \citep{pri05}, fully consistent with the \citet{sal55}
IMF. They note that the cluster IMF peaks towards smaller masses
and then declines. This is in keeping with the detailed
photometric study of \citet{sun00}, who quote an IMF slope of
$\Gamma = -1.3 \pm 0.1$, largely independent of the mass to
luminosity conversion employed. Its intrinsic velocity dispersion
is $\sigma_{\rm tot}^{\rm int} = 1.48 \pm 0.14$ mas yr$^{-1}$; at
the adopted distance, $D = 1250$ pc, this is equivalent to
$\sigma_{\rm tot}^{\rm int} \simeq 8$ km s$^{-1}$, i.e., much
closer to the dynamical state required for longer-term survival,
although still fairly large (de Grijs et al., in prep.). In the
absence of sizeable external perturbations, we speculate,
therefore, that NGC 6530 has the potential to survive for a
significant length of time, possibly of order a few $\times 10^8$
yr. However, we also note that the cluster is situated very close
to the Galactic plane in the direction of the Galactic Center,
such that external perturbations are unlikely to remain
insignificant for any length of time.

Finally, we point out that the large velocity dispersions obtained
for both young clusters imply that these objects are unlikely to
survive for any significant length of time. The velocity
dispersions of `classical' old ($\ga 10^8$ yr) bound open clusters
are of order $\la 1.5$ km s$^{-1}$ as shown by, e.g.,
\citet{loh72} in what is probably the most comprehensive study of
open cluster velocity dispersions available to date. A small
subset of these measurements have since been confirmed in more
recent studies of individual objects (de Grijs et al., in prep.).
For instance, the velocity dispersion of the $522 \pm 82$ Myr-old
\citep{pau06} central cluster in Coma Berenices is $\sigma_{\rm
1D} = 0.27 \pm 0.07$ km s$^{-1}$ (\citet{loh72}; cf. $\sigma_{\rm
1D} \sim 0.3$ km s$^{-1}$ measured in the cluster core by
\citet{ode98}); that of M67 (at an age of $\sim 4.0$ Gyr;
\citet{pau06}) ranges from $0.49 - 1.28$ km s$^{-1}$, with a best
estimate of $0.81 \pm 0.10$ km s$^{-1}$ \citep[see][de Grijs et
al., in prep.]{gir89}, while the velocity dispersion of Praesepe
(M44) is $0.46 \pm 0.2$ km s$^{-1}$ \citep{jon71}, at an age of
$753 \pm 201$ Myr \citep{pau06}. M35 ($\log( {\rm Age/yr} ) =
8.26^{+0.05}_{-0.30}$; \citet{kal03}) is characterized by
$\sigma_{\rm 1D}$ from $0.76 \pm 0.19$ km s$^{-1}$ \citep{loh72}
to $1.00 \pm 0.10$ km s$^{-1}$ \citep{mcn86}. Although we refer
the reader for further details on the kinematic structure and
long-term stability of Galactic open clusters to a forthcoming
paper (de Grijs et al., in prep.), these examples show that in
order for an open cluster to survive for a few $\times 10^8$ yr,
its velocity dispersion needs to be sufficiently small so as to
prevent early dissolution of the cluster due to internal kinematic
effects. Neither of the clusters discussed in this paper satisfies
this empirical condition. The only old ($\log( {\rm Age/yr} ) =
8.4 \pm 0.1$; \citet{sun99}) Galactic open cluster with a measured
velocity dispersion in excess of the range indicated above is M11,
with $\sigma_{\rm 1D}$ from $1.21 \pm 0.35$ km s$^{-1}$
\citep{mat84} to 2.9 km s$^{-1}$ \citep{mcn77}; the latter
measurement was later reduced to 2.0 km s$^{-1}$ \citep{mcn86}.
This high velocity dispersion, for an ostensibly stable {\it
massive} open cluster, is still more than a factor of two below
that of NGC 6530, which is both significantly less massive and
less compact than M11, and thus less stable to kinematic
dissolution.

The main points of the present study can be summarized as follows:

\begin{enumerate}
\item From the photographic plate data of the Shanghai Astronomical
Observatory, the proper motions of stars in the region of the open
clusters NGC 2244 and NGC 6530 were reduced by means of a
central-overlapping technique, and the distribution parameters of the
clusters as well as the membership probabilities of the individual
stars in the cluster region were determined using a maximum likelihood
principle.

\item Both clusters show clear evidence of mass segregation, but there
is no definite evidence for a velocity (dispersion)-mass (or,
equivalently, a velocity-luminosity) dependence in these clusters. The
observed mass segregation might be due to a combination of both
initial conditions and relaxation processes.

\item Based on arguments related to the clusters' IMFs, in conjunction
with our new measurements of their internal velocity dispersions, we
provide strong arguments in favor of the dissolution of NGC 2244 on
very short time-scales, while we speculate that NGC 6530 may be more
stable against the effects of internal two-body relaxation.

\end{enumerate}

\acknowledgments

RdG was partially supported by an ``International Joint Project''
grant, jointly funded by the Royal Society in the UK and the
Chinese National Science Foundation (NSFC). LC \& JLZ want to
thank Dr. Peter Stetson of DAO for his kind help in measuring
plates with the PDS machine. LC and JLZ are supported by the NSFC
(Grant Nos. 10333050, 10333020).This research has made use of the
WEBDA database, operated at the Institute for Astronomy of the
University of Vienna.

\clearpage

\begin{figure}
\plottwo{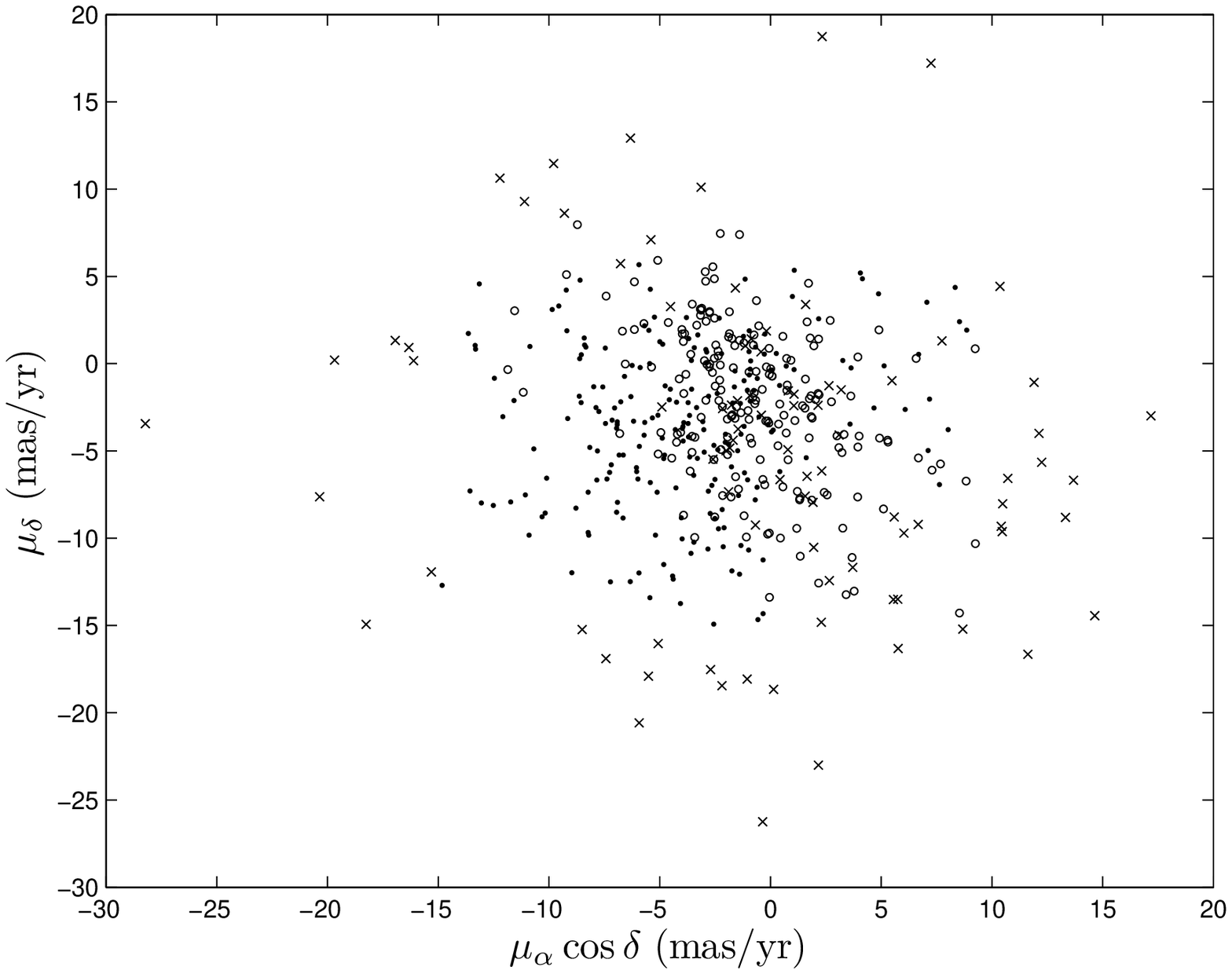}{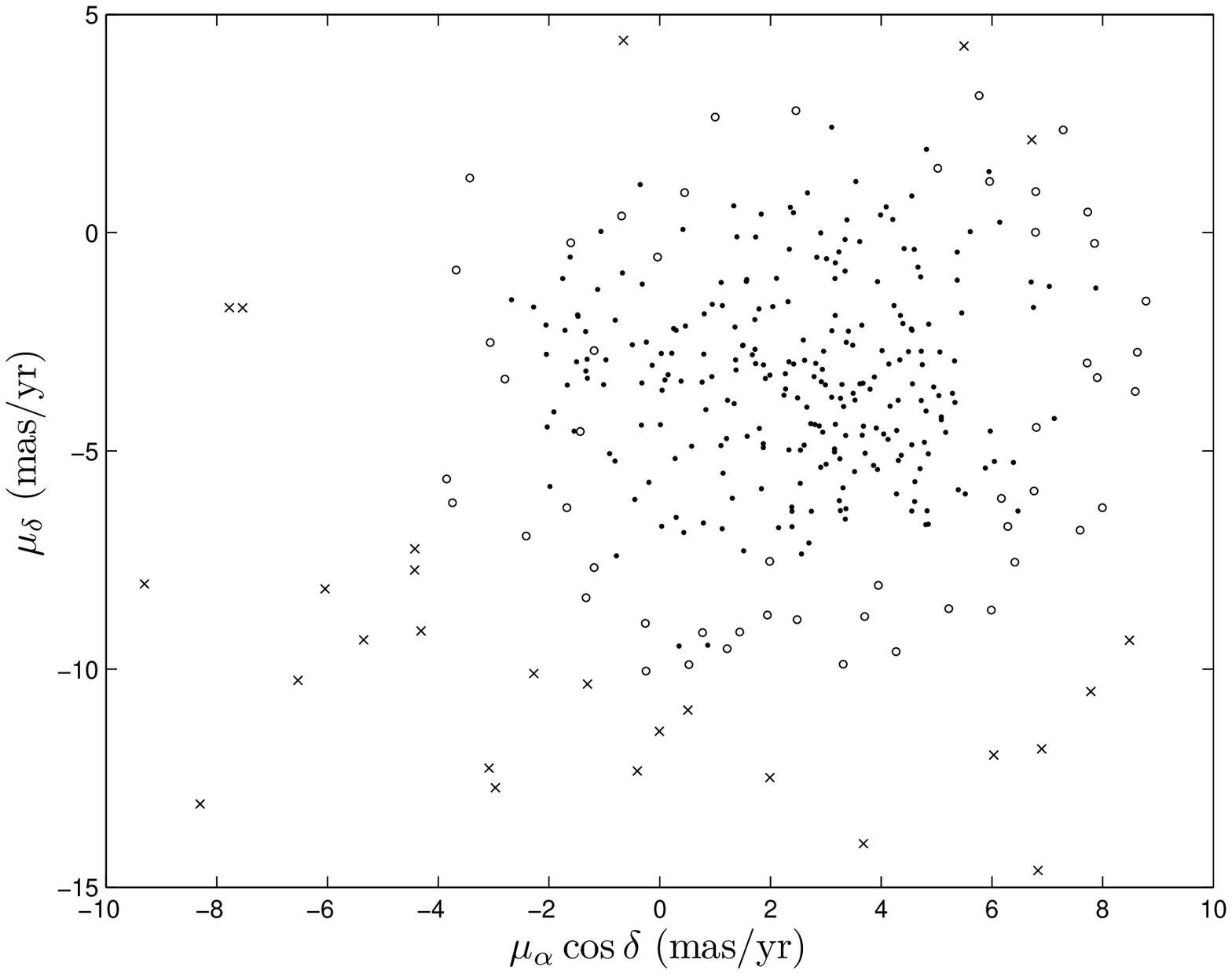} \caption{Vector-point diagrams for both
NGC 2244 (left) and NGC 6530 (right); the solid points represent
the stars with membership probabilities $p \ge 0.9$, whereas the
crosses indicate the $p \le 0.5$. The open circles have $p$ values
in between these boundaries.\label{fig1}}
\end{figure}

\begin{figure}
\plottwo{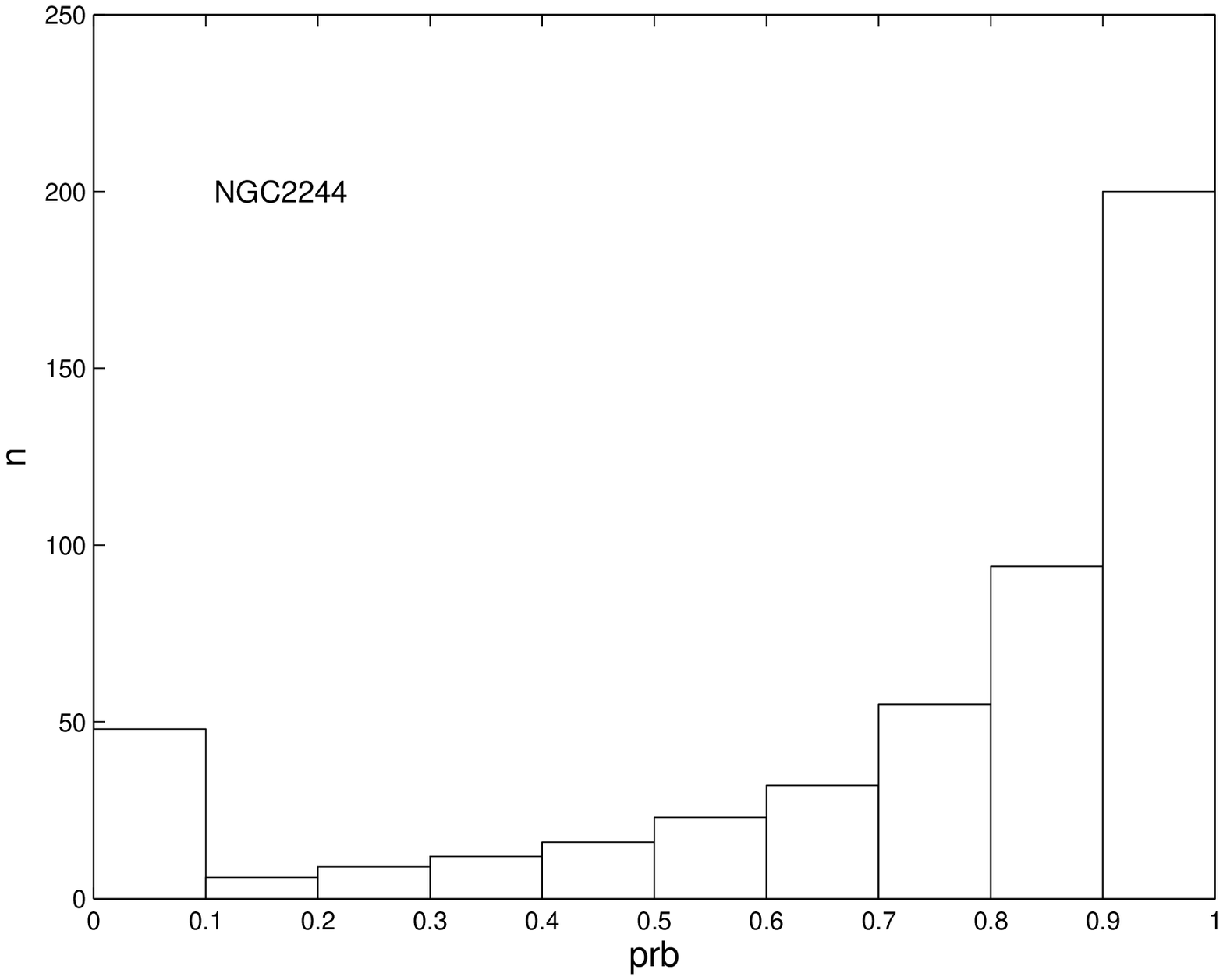}{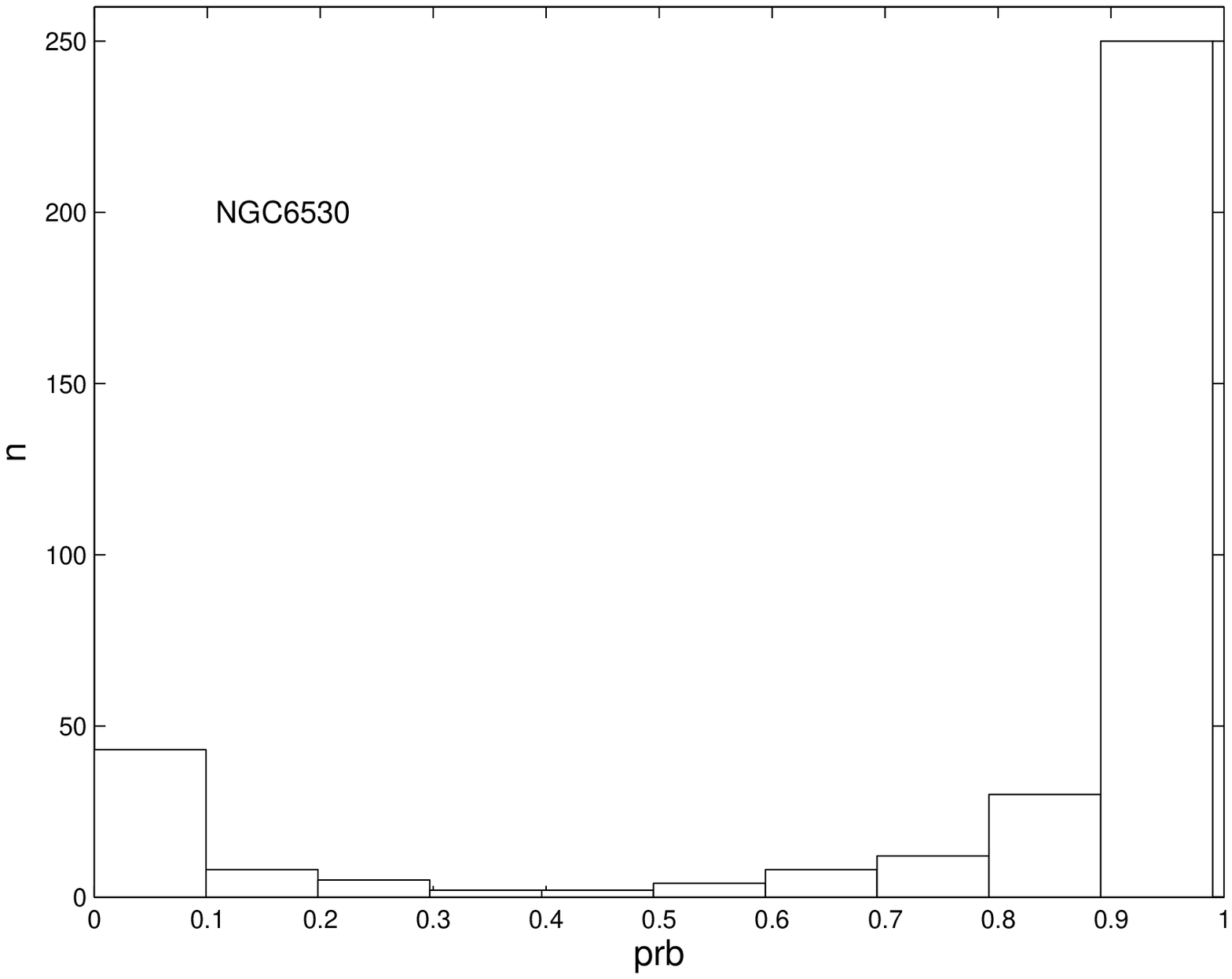} \caption{Histogram of the stellar
membership probabilities. {\it Left panel:} 495 stars in the
region of NGC 2244; {\it Right panel:} 364 stars in region of NGC
6530.\label{fig2}}
\end{figure}

\begin{figure}
\plottwo{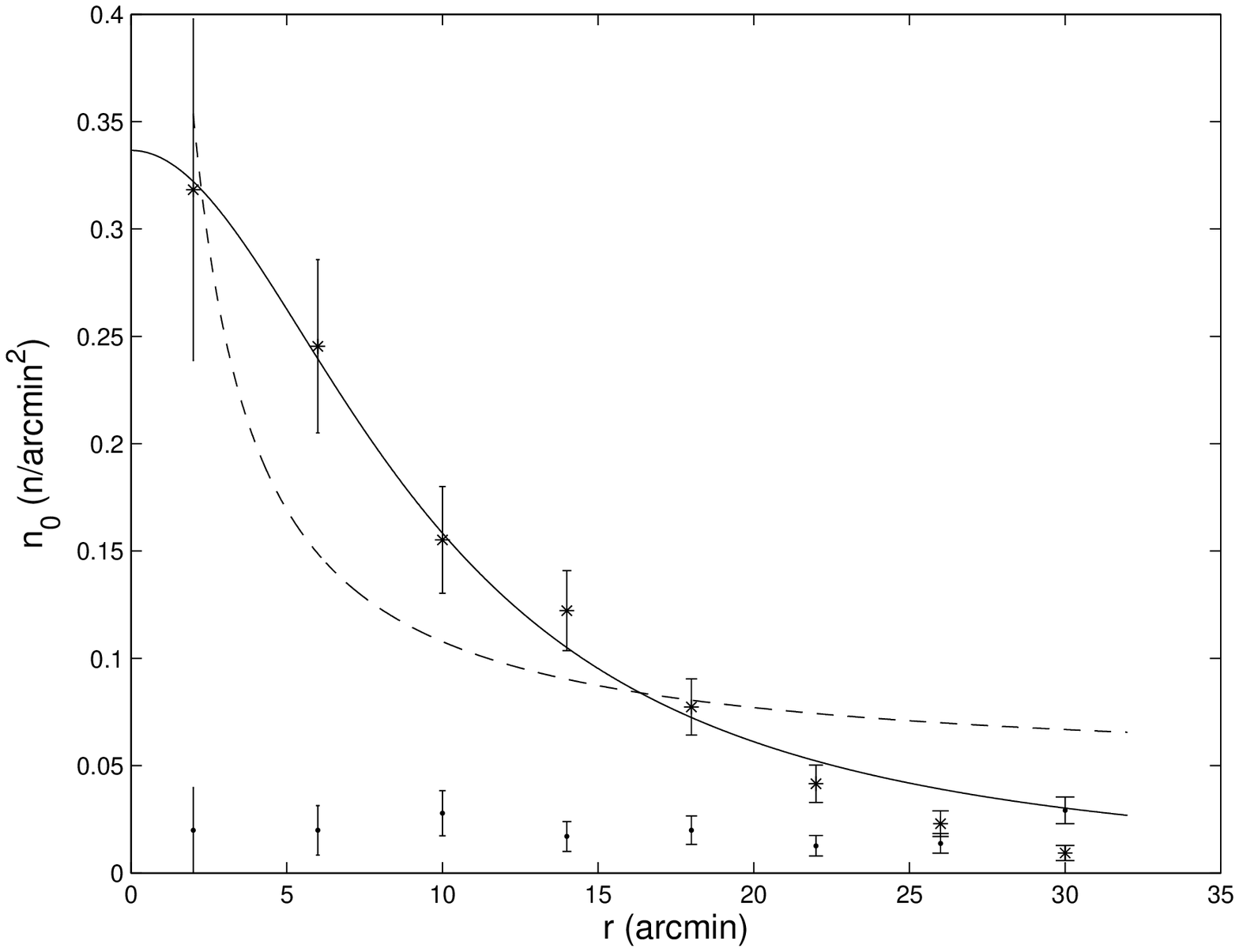}{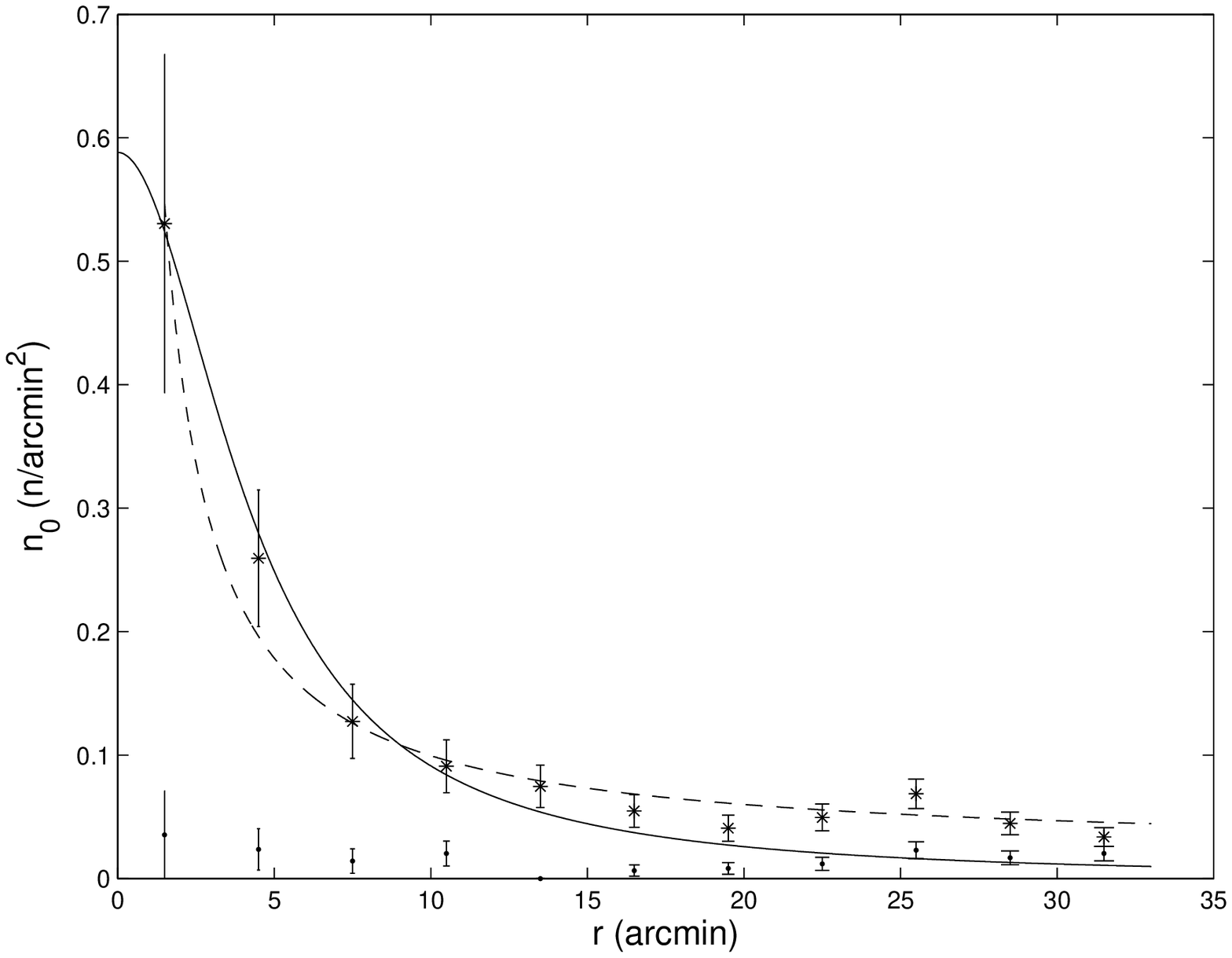} \caption{Radial number density profiles
of the stars in the open cluster regions. {\it Left:} NGC 2244;
{\it Right:} NGC 6530. $\ast$: field stars with membership
probabilities $p\leq0.5$; $\cdot$: member stars with $p\geq0.9$.
Solid lines: Gaussian profiles; dashed lines: $R^{-1}$ profiles.
\label{fig3}}
\end{figure}

\begin{figure}
\plottwo{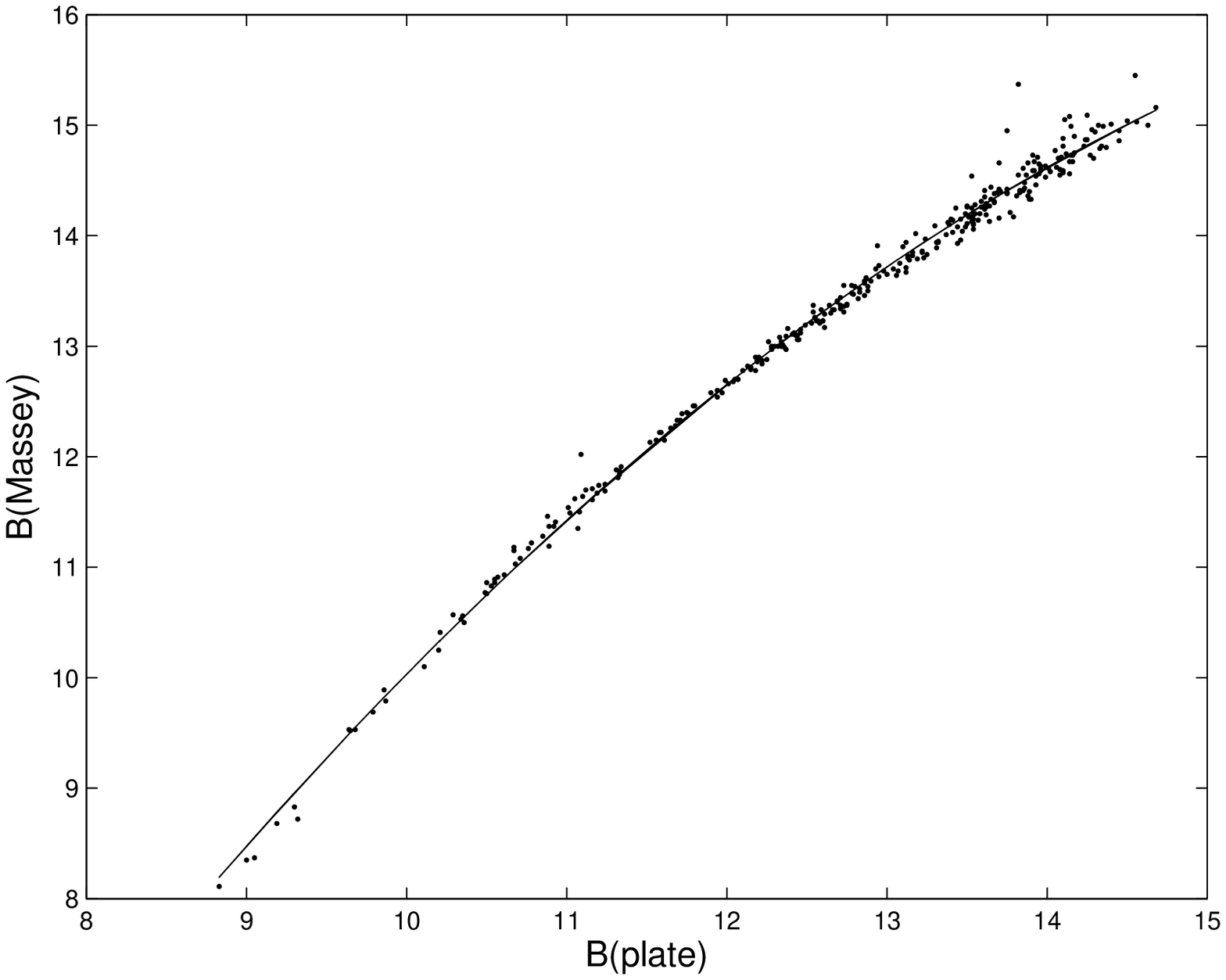}{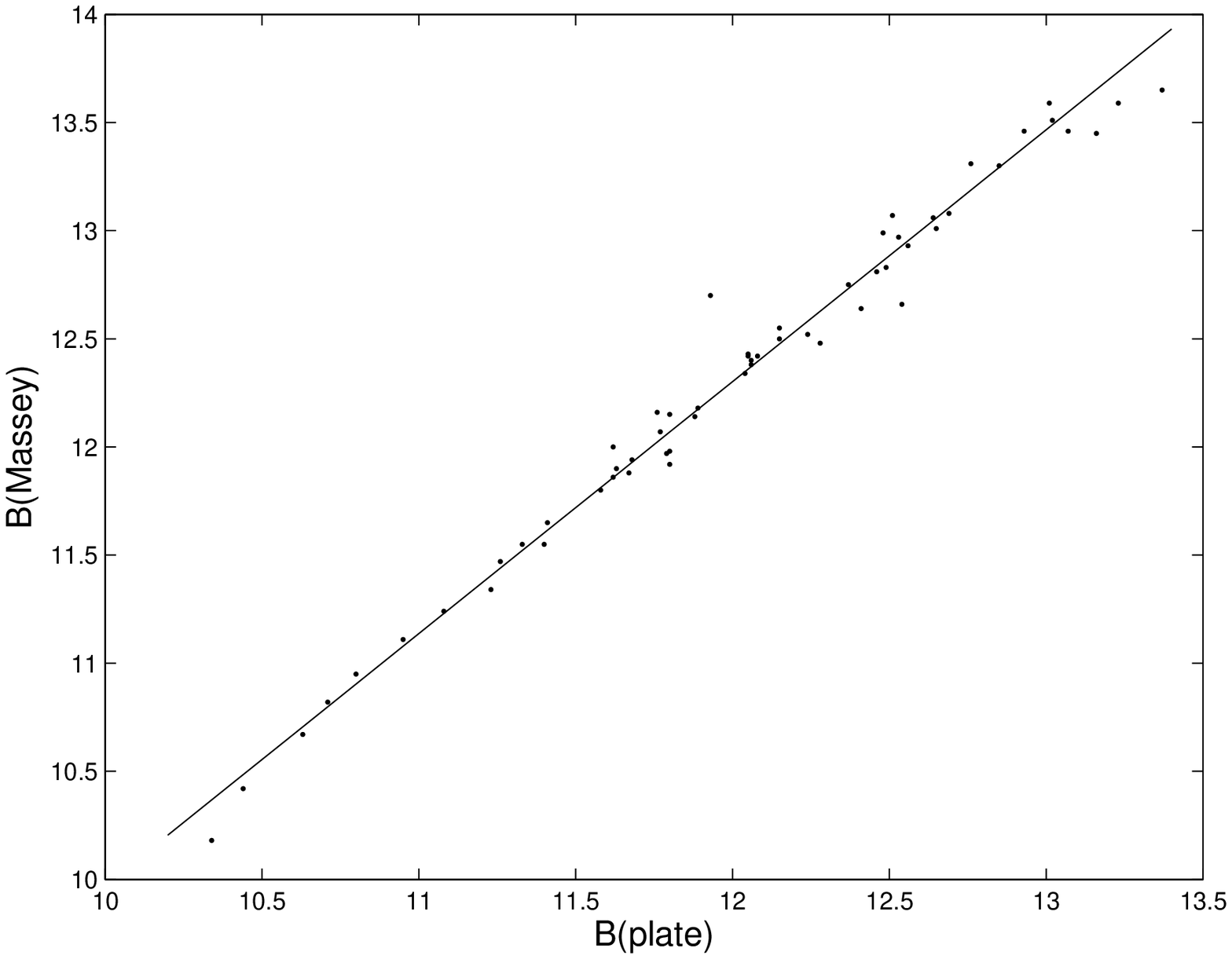} \caption{Comparison between our
instrumental $B_p$ photometry and the $B$ magnitudes of the $UBV$
system. {\it Left:} 323 stars in common with Massey et al. (1995);
{\it Right:} 58 stars in common with \citet{pri05} \label{fig4}}
\end{figure}

\begin{figure}
\epsscale{.80} \plotone{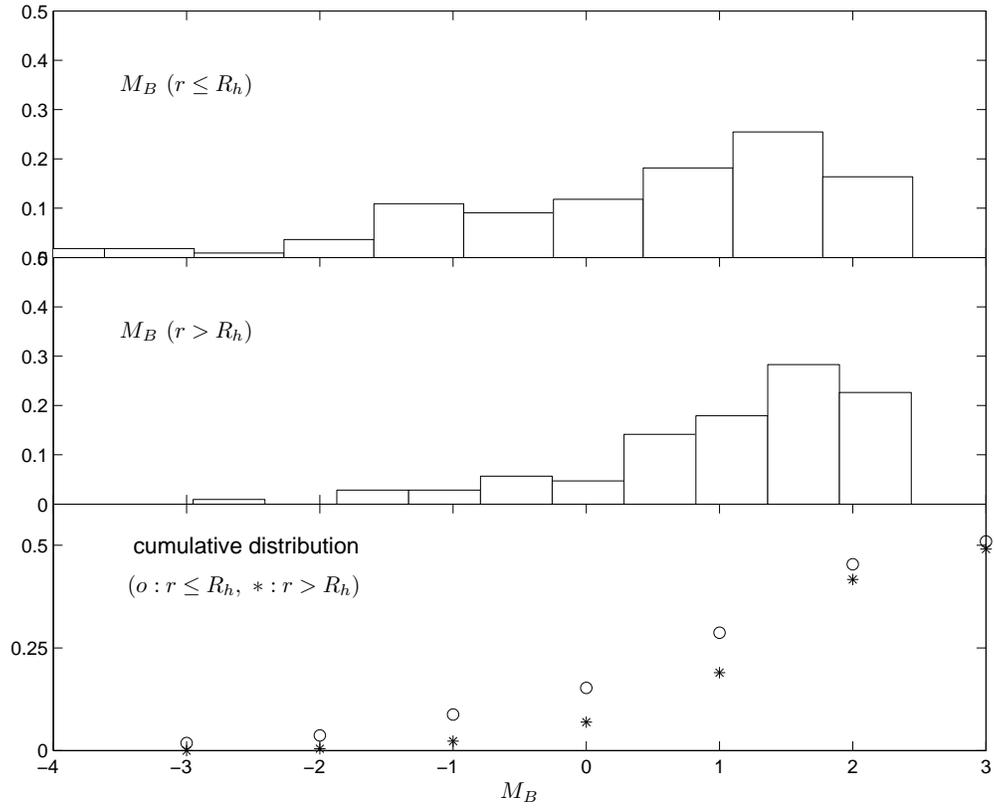} \caption{Luminosity function of NGC
2244. {\it Top:} 110 member stars in the inner area $(r\leq 14')$;
{\it middle:} 106 member stars in the outer region $(r> 14')$; {\it
bottom:} cumulative luminosity functions for both radial regions
\label{fig5}}
\end{figure}

\begin{figure}
\epsscale{.80} \plotone{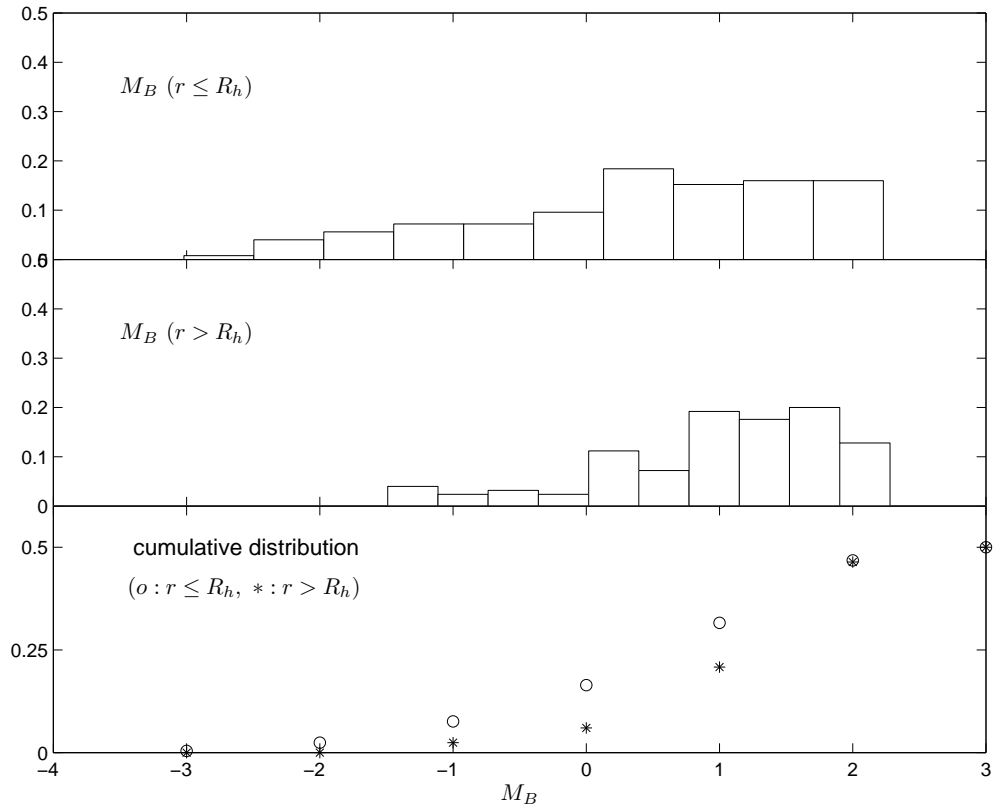} \caption{Luminosity function of NGC
6530. {\it Top:} 125 member stars in the inner area $(r\leq 21')$;
{\it middle:} 125 member stars in the outer region $(r> 21')$; {\it
bottom:} cumulative luminosity functions for both radial regions
\label{fig6}}
\end{figure}

\begin{figure}
\epsscale{.80} \plotone{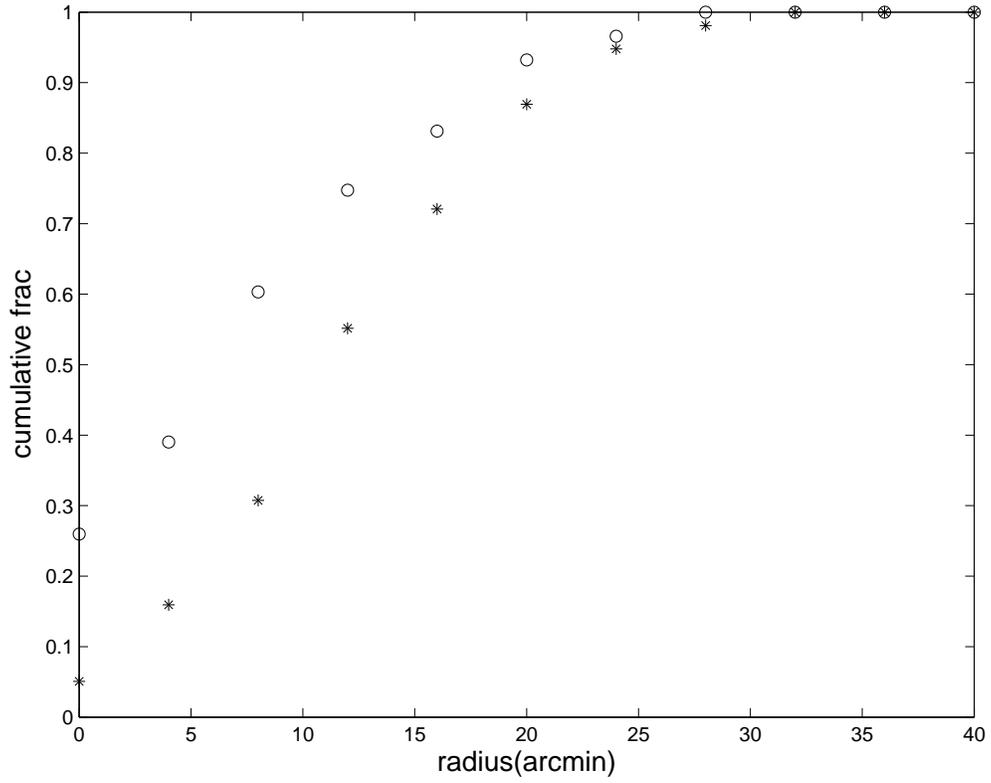} \caption{Normalized cumulative
radial number density profile for NGC 2244 members with $m_B \leq
13$ (o) and $m_B > 13$ ($\ast$) \label{fig7}}
\end{figure}

\begin{figure}
\epsscale{.80} \plotone{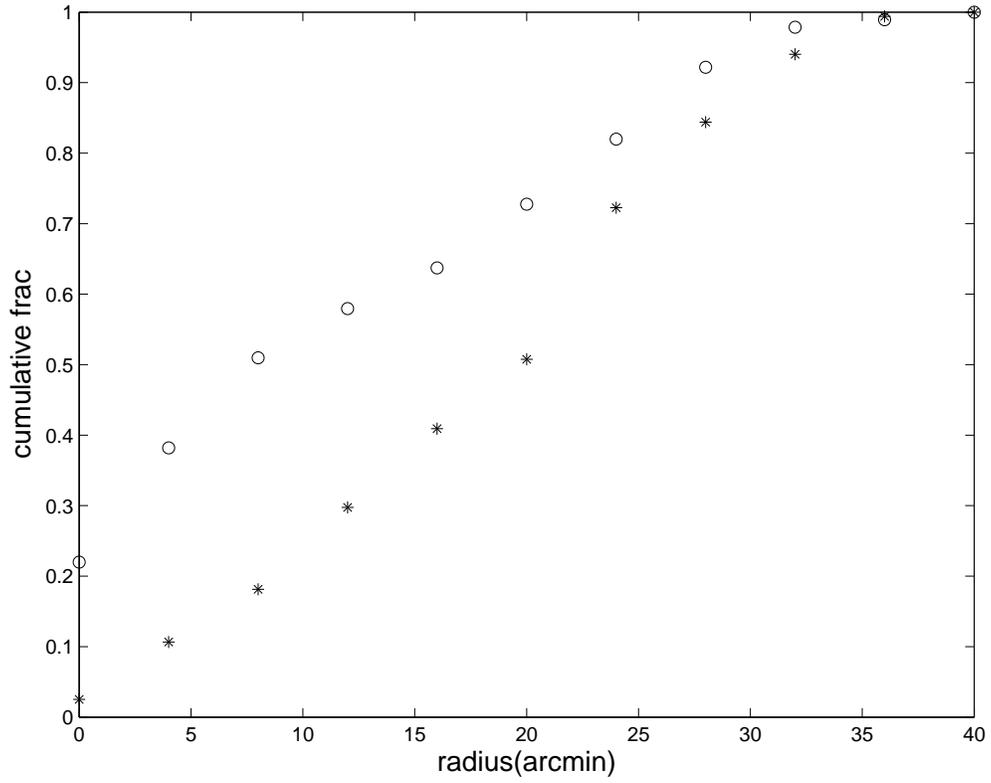} \caption{Normalized cumulative
radial number density profile for NGC 6530 members with $m_B \leq
12$ (o) and $m_B > 12$ ($\ast$) \label{fig8}}
\end{figure}

\begin{figure}
\epsscale{.80} \plotone{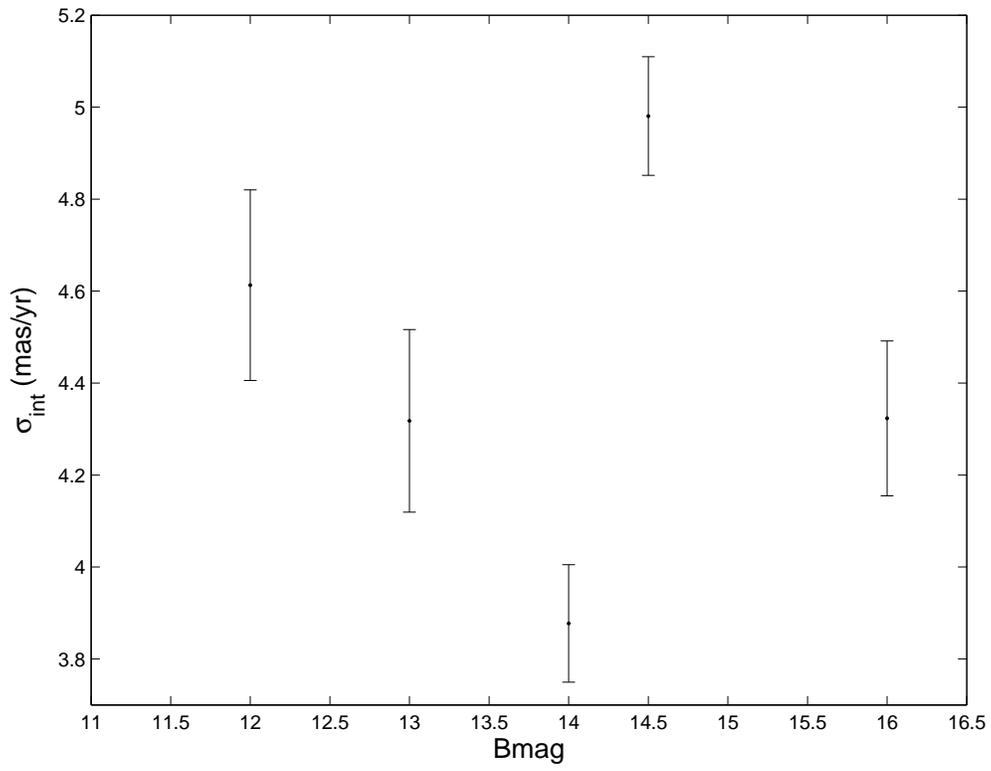} \caption{Proper motion dispersion
of NGC 2244 members as a function of $m_B$ magnitude.
\label{fig9}}
\end{figure}

\begin{figure}
\epsscale{.80} \plotone{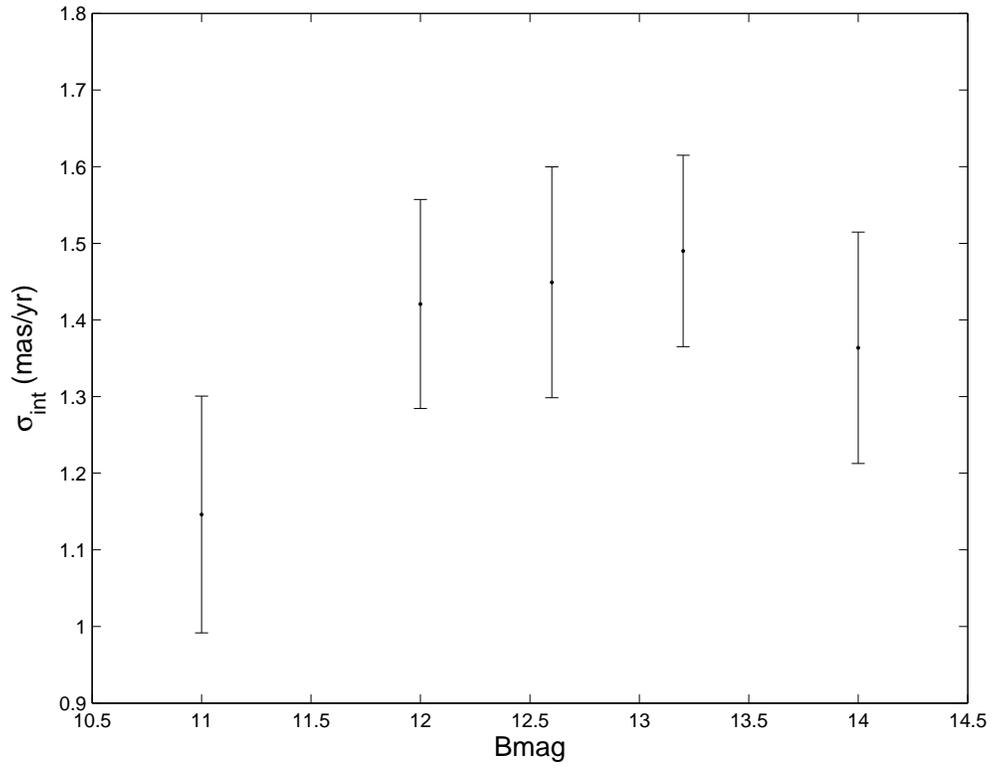} \caption{Proper motion dispersion
of NGC 6530 members as a function of $m_B$ magnitude.
\label{fig10}}
\end{figure}

\clearpage

\begin{deluxetable}{cccccc}
\label{tab1}
\tablecolumns{6} \tablewidth{0pc} \tablecaption{Internal levels of uncertainty on the stellar positions
 and proper motions} \tablehead{ \colhead{}    & \multicolumn{2}{c}{NGC 2244} & \colhead{} &
\multicolumn{2}{c}{NGC 6530} \\
\cline{2-3} \cline{5-6} \\
\colhead{Parameter} & \colhead{Median error}   & \colhead{Maximum
error} & \colhead{}    & \colhead{Median error}   &
\colhead{Maximum error} }
\startdata
$\alpha$  & $0^{\rm s}.003$ & $0^{\rm s}.017$ & & $0^{\rm s}.003$& $0^{\rm s}.014$ \\
$\delta$  & $0''.043$ & $0''.33$ & & $0''.06$& $0''.16$ \\
$\mu_{\alpha}\cos\delta$  & 1.39 mas yr$^{-1}$ & 8.88 mas yr$^{-1}$ & & 0.95 mas yr$^{-1}$ & 2.74 mas yr$^{-1}$ \\
$\mu_{\delta}$  & 1.56 mas yr$^{-1}$ & 9.54 mas yr$^{-1}$ & & 1.21 mas yr$^{-1}$ & 4.05 mas yr$^{-1}$ \\
\enddata
\end{deluxetable}

\clearpage

\begin{deluxetable}{cccc}
\tablecolumns{4} \tablewidth{0pc} \tablecaption{Estimates of
the distribution parameters} \tablehead{ \colhead{}    &
\colhead{NGC 2244} & \colhead{} &
\colhead{NGC 6530} \\
\cline{2-2} \cline{4-4} \\
\colhead{Parameter} & \colhead{Estimate and uncertainty}   &
\colhead{}    & \colhead{Estimate and uncertainty}    } \startdata
$\mu_{x{\rm c}}$(mas yr$^{-1}$)  & $-2.51\pm0.23$  & & $2.59\pm0.16$ \\
$\mu_{y{\rm c}}$(mas yr$^{-1}$)  & $-3.02\pm0.25$  & & $-3.48\pm0.17$  \\
$\mu_{x{\rm f}}$(mas yr$^{-1}$)  & $-0.41\pm0.38$  & & $3.62\pm0.34$ \\
$\mu_{y{\rm f}}$(mas yr$^{-1}$)  & $-2.81\pm0.28$  & & $-7.69\pm0.88$  \\
$\sigma_{0} $(mas yr$^{-1}$)  & $1.48\pm0.14$  & & $2.41\pm0.40$ \\
$\sigma_{x0}$(mas yr$^{-1}$)  & $2.99\pm0.25$  & & $9.19\pm0.17$  \\
$\sigma_{y0}$(mas yr$^{-1}$)  & $4.81\pm0.23$  & & $8.71\pm0.36$ \\
$\rho       $          & $-0.59\pm0.04$  & & $0.22\pm0.07$  \\
$\alpha(')$            & $13.24\pm0.14$  & & $24.45\pm0.70$ \\
$   g       $          & $19.56\pm0.79$  & & $8.84\pm0.21$  \\
\enddata
\end{deluxetable}

\clearpage


\begin{deluxetable}{rrrrrrrrr}
\tabletypesize{\scriptsize} \rotate \tablecaption{Proper motions,
membership probabilities and $B$ magnitude of 495 stars in the
region of NGC 2244} \tablewidth{0pt} \tablehead{ \colhead{No.} &
\colhead{R.A. (J2000.0)} & \colhead{Dec (J2000.0)} &
\colhead{$\mu_{\alpha}\cos\delta$} & \colhead{$\sigma_{\rm PA}$} &
\colhead{$\mu_{\delta}$} &
\colhead{$\sigma_{\rm PD}$} & \colhead{$B$ (mag)} & \colhead{$p$} \\
\colhead{} & \colhead{} & \colhead{} & \colhead{mas yr$^{-1}$} &
\colhead{mas yr$^{-1}$} & \colhead{mas yr$^{-1}$} & \colhead{mas yr$^{-1}$} &
\colhead{} & \colhead{}
 }
\startdata
   1 & 06 31 35.826 & +04 30 51.70 & $-$6.142 & 0.919 &    4.685 & 0.405 & 8.720 & 0.62  \\
   2 & 06 32 10.470 & +04 57 59.75 & $-$1.600 & 1.486 & $-$0.421 & 2.184 & 8.350 & 0.92  \\
   3 & 06 31 36.322 & +04 55 59.53 &    0.336 & 0.464 & $-$4.710 & 1.008 & 8.830 & 0.89  \\
   4 & 06 31 38.396 & +05 01 36.38 & $-$1.673 & 0.639 & $-$0.232 & 1.303 & 8.110 & 0.89  \\
   5 & 06 31 16.839 & +04 37 25.92 & $-$0.074 & 0.425 & $-$3.398 & 1.127 & 9.039 & 0.73  \\

\enddata
\tablecomments{This table is published in its entirety in the
electronic edition.  A portion is shown here for guidance
regarding its form and content.}
\end{deluxetable}

\clearpage

\begin{deluxetable}{rrrrrrrrr}
\tabletypesize{\scriptsize} \rotate \tablecaption{Proper motions,
membership probabilities and $B$ magnitude of 364 stars in the
region of NGC 6530} \tablewidth{0pt} \tablehead{ \colhead{No} &
\colhead{R.A. (J2000.0)} & \colhead{Dec (J2000.0)} &
\colhead{$\mu_{\alpha}\cos\delta$} & \colhead{$\sigma_{\rm PA}$} &
\colhead{$\mu_{\delta}$} &
\colhead{$\sigma_{\rm PD}$} & \colhead{$B$ (mag)} & \colhead{$p$} \\
\colhead{} & \colhead{} & \colhead{} & \colhead{mas yr$^{-1}$} &
\colhead{mas yr$^{-1}$} & \colhead{mas yr$^{-1}$} & \colhead{mas yr$^{-1}$} &
\colhead{} & \colhead{}
 }
\startdata
   1 & 18 04 25.848 & $-$24 23 08.27 & 6.469 & 0.815 & $-$6.376 &  2.249 &   8.55 &
   0.94 \\
   2 & 18 06 04.702 & $-$24 11 43.84 & 5.989 & 0.591 & $-$8.649 &  1.670 &   8.64 &
   0.83 \\
   3 & 18 04 15.222 & $-$24 11 00.07 & 2.958 & 0.877 & $-$2.713 &  1.067 &   9.12 &
   0.99 \\
   4 & 18 02 39.798 & $-$24 14 47.40 & 7.732 & 0.405 &    0.474 &  3.538 &   9.22 &
   0.74 \\
   5 & 18 04 14.538 & $-$24 14 36.31 & 8.511 & 0.853 &    9.241 &  1.665 &   9.35 &
   0.00 \\

\enddata
\tablecomments{This table is published in its entirety in the
electronic edition.  A portion is shown here for guidance
regarding its form and content.}
\end{deluxetable}

\clearpage
\begin{deluxetable}{cccccc}
\tablecolumns{6} \tablewidth{0pc} \tablecaption{Half-number radii
of cluster members for different magnitude ranges} \tablehead{
\colhead{} & \multicolumn{2}{c}{NGC 2244} & \colhead{} &
\multicolumn{2}{c}{NGC 6530} \\
\cline{2-3} \cline{5-6} \\
\colhead{$B$} & \colhead{$N$ (stars)}   & \colhead{Half-number
radius} & \colhead{} & \colhead{$N$ (stars)}   &
\colhead{Half-number radius} } \startdata
All member stars  & 216 & $13.5'$ & & 250 & $20.7'$ \\
$M_{B}\leq0$ mag  &  38 &  $9.2'$ & &  62 &  $9.8'$ \\
$M_{B}>0$ mag     & 178 & $14.3'$ & & 188 & $23.2'$ \\
\enddata
\end{deluxetable}

\end{document}